\documentclass{article}

\usepackage{arxiv}

\usepackage[T1]{fontenc}    
\usepackage{hyperref}       
\usepackage{url}            
\usepackage{booktabs}       
\usepackage{amsfonts}       
\usepackage{nicefrac}       
\usepackage{microtype}      
\usepackage{lipsum}
\usepackage{graphicx}
\graphicspath{ {./images/} }
\usepackage{epstopdf, epsfig}
\usepackage{caption}
\usepackage{color, soul}
\usepackage[colorinlistoftodos]{todonotes}

\usepackage{longtable} 
\usepackage{multirow} 
\usepackage{amsmath} 
\usepackage[T1]{fontenc} 
\usepackage{tabularx} 
\usepackage{changepage} 
\usepackage[numbers]{natbib} 
\usepackage{float} 
\usepackage[running]{lineno}
\usepackage{xcolor}

\usepackage{relsize}
\usepackage{url}
\usepackage{appendix} 

\renewcommand{\figurename}{FIG.}
\usepackage{hyperref}

\title{\rm{Unravelling the turbulent structures of temperature variations during a wind gust event: a case study}}

\author{
Subharthi Chowdhuri \\
  Indian Institute of Tropical Meteorology\\
  Ministry of Earth Sciences, India\\
  Dr. Homi Bhaba Road, Pashan, Pune-411008\\
  Email: \href{mailto:subharthi.cat@tropmet.res.in}{subharthi.cat@tropmet.res.in}
   \And
Kiran Todekar, Palani Murugavel \\
  Indian Institute of Tropical Meteorology\\
  Ministry of Earth Sciences, India\\
  Dr. Homi Bhaba Road, Pashan, Pune-411008\\
  \And
Anand K Karipot \\
  Savitribai Phule Pune University\\
  Ganeshkhind, Pune-411007\\
  Pune, India\\
    \And
Thara V Prabha \\
  Indian Institute of Tropical Meteorology\\
  Ministry of Earth Sciences, India\\
  Dr. Homi Bhaba Road, Pashan, Pune-411008\\
}

\begin{document}
\maketitle
\begin{abstract}
The simultaneous observations from a Doppler weather radar and an instrumented micrometeorological tower, offer an opportunity to dissect the effects of a gust front on the surface layer turbulence in a tropical convective boundary layer. We present a case study where a sudden drop in temperature was noted at heights within the surface layer during the passage of a gust front in the afternoon time. Consequently, this temperature drop created an interface which separated two different turbulent regimes. In one regime the turbulent temperature fluctuations were large and energetic, whereas in the other regime they were weak and quiescent. Given its uniqueness, we investigated the size distribution and aggregation properties of the turbulent structures related to these two regimes. We found that, the size distributions of the turbulent structures for both of these regimes displayed a clear power-law signature. Since power-laws are synonymous with scale-invariance, this indicated the passing of the gust front initiated a scale-free response which governed the turbulent characteristics of the temperature fluctuations. We propose a hypothesis to link such behaviour with the self organized criticality as observed in the complex systems. However, the temporal organization of the turbulent structures, as indicated by their clustering tendencies, differed between these two regimes. For the regime, corresponding to large temperature fluctuations, the turbulent structures were significantly clustered, whose clustering properties changed with height. Contrarily, for the other regime where the temperature fluctuations were weak, the turbulent structures remained less clustered with no discernible change being observed with height.  
\end{abstract}

\keywords{Clustering \and Convective surface layer \and Gust front \and Size distribution \and Turbulent structures}

\section{Introduction}
\label{intro}
Gust fronts are a typical phenomena associated with convective outflows from thunderstorms that are growing in a dry environment \citep{droegemeier1985three,mueller1987dynamics,weckwerth1992initiation}. The evaporative cooling of rain and the associated downdrafts initiate the cold pool, which introduces cold and moist air into the dry and sub-saturated boundary layer. As the moist air interfaces with the dry surrounding air, the density differences between the two air masses create the gust front. The speed of the wind gusts may attain values more than 15 m s$^{-1}$ associated with temperature drops more than 6 to 10 K due to the intrusion of the cold air (see \citet{hutson2017using} and references therein). The gust front propagates as the cold pool expands and sometimes secondary gust fronts are also generated by the storm outflows \citep{doviak1984atmospheric}. There can also be multiple gust fronts from other storms in the neighbourhood. Gust fronts play an important role in the formation of new convection and the organization of convection \citep{kingsmill1995convection}. The understanding of the gust front and their interaction with the surface is of importance for various applications such as the nowcasting, pollution dispersion, wind damage to structures, etc. 

The convective outflows and gust fronts can influence the surface fluxes and turbulence in the boundary layer \citep{oliveira2020planetary}. The signatures of the gust fronts are imprinted on the surface measurements in the form of a sudden increase in the wind speed followed by a sharp drop in the temperature. This typical gust front feature has remarkable resemblance with the temperature micro-fronts, commonly observed and extensively well-studied in the context of a nocturnal boundary layer over the mid latitudes \citep{blumen1999analysis,prabha2007low,mahrt2019microfronts}. However, to the best of our knowledge there are no systematic investigations available about the effect of wind gust on the turbulent characteristics of temperature fluctuations in the surface layer of a tropical convective boundary layer.

The surface layer is a generalization of the inertial layer of unstratified wall-bounded flows by including the effect of buoyancy, where the effects of surface roughness are no longer important and the modulations by the outer eddies are not too strong \citep{barenblatt1996scaling,davidson2015turbulence}. In a convective surface layer flow, the temperature fluctuations behave like an active scalar, where they preferentially couple with the velocity fluctuations in the vertical direction to transport heat and drive the flow \citep{tennekes1972first}. Along with that, the turbulent fluctuations in temperature also share characteristics akin to the laboratory Rayleigh B\'{e}nard convection experiments \citep{adrian1986turbulent,balachandar1991probability,wang2019turbulent}. The large-eddy simulations of \citet{khanna1998three} and aircraft experiments of \citet{williams1992composite,williams1993interactions} demonstrate that similar cellular convective structures as found in the laboratory experiments also exist in a convective surface layer. These studies illustrate that the positive temperature fluctuations aggregate along the cell edges, whereas the negative fluctuations reside within the cell center. Apart from that, in a convective surface layer the temperature fluctuations are also unevenly distributed due to their non-Gaussian nature in the local free-convection limit. This uneven distribution is caused due to the intermittent bursting of the warm plumes rising from the ground, interspersed with relatively more frequent quiescent cold plumes bringing well-mixed air from aloft \citep{chu1996probability,liu2011probability,garai2013interaction,lyu2018high}. As a consequence, the temperature fluctuations in a convective surface layer display intermittent characteristics. 

Intermittency is defined as a property of a signal which is quiescent for much of the time and occasionally burst into life with unexpectedly high values more common than in a Gaussian signal \citep{davidson2015turbulence}. In general, there are two aspects which characterize an intermittent turbulent signal \citep{bershadskii2004clusterization,sreenivasan2006clustering,cava2019submeso}. The first one of these is related to an uneven distribution of quiescent versus energetic states in space or time, caused due to the aggregation or clustering of the turbulent structures. On the other hand, the second one is related to the amplitude variability associated with the signal. 

However, in a convective surface layer, systematic investigation of the intermittent characteristics of temperature fluctuations by separating the clustering and amplitude variability effects is quite rare. In fact, there are only two available studies by \citet{cava2009effects} and \citet{cava2012role}, where such an analysis has been carried out. By employing a telegraphic approximation (TA) of the temperature signal, they showed that the temperature fluctuations displayed significant clustering due to the presence of the coherent ramp-cliff patterns. Nevertheless, they also discovered that the clustering effect in temperature was significantly dependent on the surface heterogeneity, such that the temperature was more clustered within a canopy sub layer rather than over a bare soil. They attributed this difference to a hypothesis that the large-scale structures such as ramps are less distorted in the canopy sub layers when compared to their atmospheric counterparts. 

Therefore, it is prudent to ask in a convective surface layer,
\begin{enumerate}
    \item How does the presence of a gust front modify the aggregation (clustering) properties of the turbulent structures related to the temperature fluctuations?
    \item Does this modification occur across all the turbulent scales or is there any particular threshold beyond which the effect ceases to exist?
\end{enumerate}
These are the primary research questions which motivate this study. In our course of investigation, we employ novel statistical methods \citep{bershadskii2004clusterization,poggi2009flume,chowdhuri2020persistence} to carry out an in-depth analysis of the turbulent structures of temperature variations associated with a gust front event which occurred during the afternoon time of 22-September-2018. The reason this day has been chosen for our analysis is because the availability of simultaneous and co-located observations from a Doppler weather radar and from a micrometeorological tower. Together these measurements present a unique opportunity to scrutinize the spatial morphology of the wind gust as it approached the location of the tower and dissect the associated effects on the turbulent temperature characteristics. 

The present article is organized in three different sections. In Sect. \ref{Data}, we describe the dataset, in Sect. \ref{results} we present and discuss the results, and lastly in Sect. \ref{concl} we conclude our findings and present future research direction. 

\section{Dataset description}
\label{Data}
To investigate the turbulent characteristics, we have used the micrometeorological dataset from a 50-m instrumented tower erected over a non-irrigated grassland in Solapur, India (17.6$^{\circ}$ N, 75.9$^{\circ}$ E, 510 m above mean sea level). This dataset was collected during the Integrated Ground Observational Campaign (IGOC), as a part of the fourth phase (2018-2019) of the Cloud Aerosol Interaction and Precipitation Enhancement Experiment (CAIPEEX). The fourth phase of the CAIPEEX was conducted over the arid rain shadow region of Western-Ghats with a dense observational network. Further details about the CAIPEEX program and its objectives can be found in \citet{kulkarni2012cloud} and \citet{prabha2011microphysics}.    

\begin{figure*}[h]
\centering
\resizebox{.6\textwidth}{0.8\textwidth}{\includegraphics {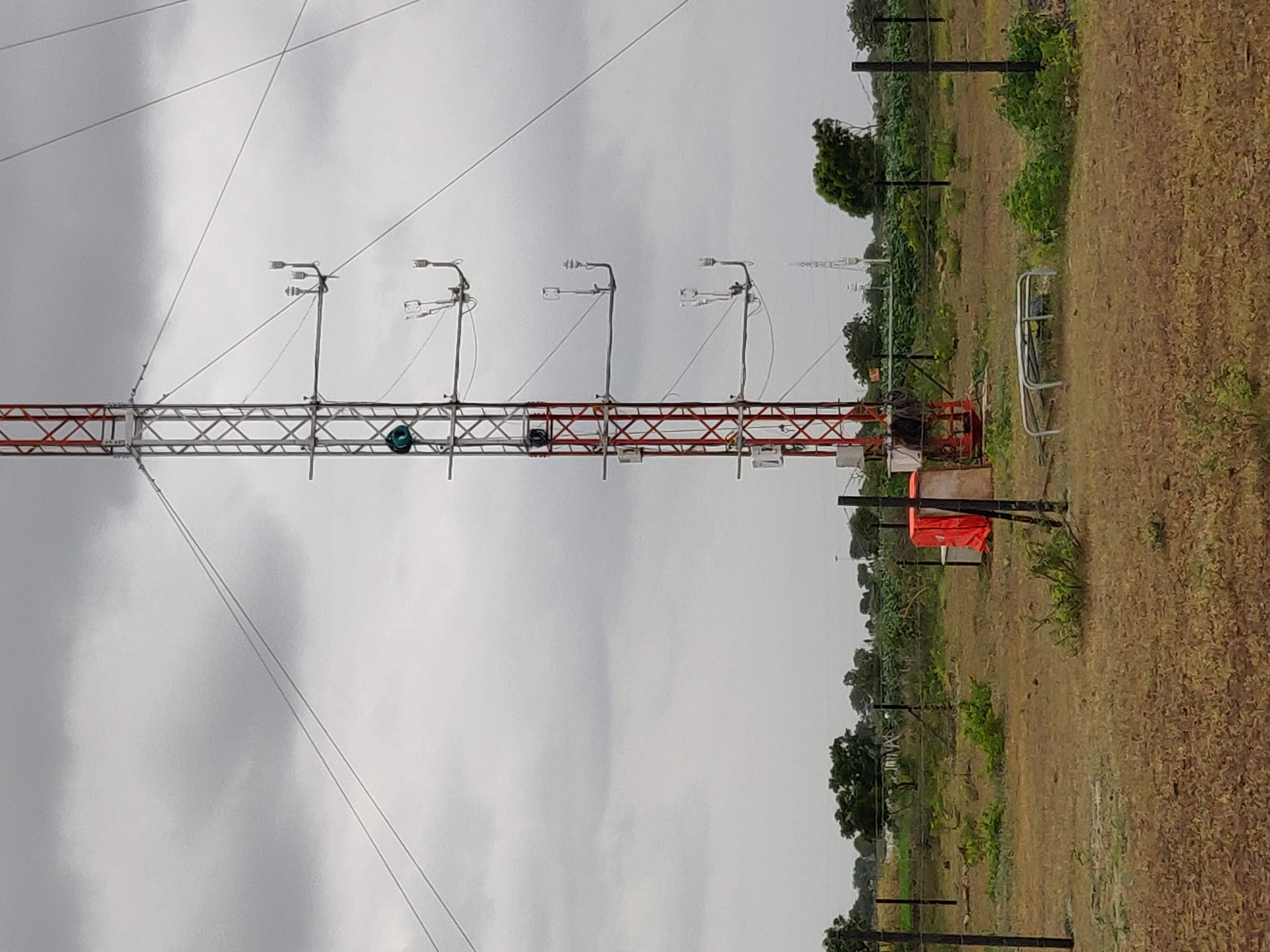}}
\vspace{3mm}
  \caption{A partial view of the 50-m instrumented tower, looking towards the South. The booms where the instruments are mounted face towards the West. The fetch area at the South-West sector of the tower includes scattered trees and bushes, representing a typical site.}
\label{fig:1}
\end{figure*}

Figure \ref{fig:1} shows a partial view of the 50-m tower, along with its surroundings. The terrain was relatively flat at the tower location, populated with grassland, scattered trees and bushes at the South-West sector, having an average roughness height of about 10 cm in the southerly direction. Note that, a small one-storeyed building was located at the North side (not seen in Fig. \ref{fig:1}) of the tower, approximately at a distance of 100 m. Due to this reason, the data were only used when the wind direction was from the South-West sector.

The tower was instrumented with the sonic anemometers (Gill Windmaster-Pro) at four levels, corresponding to the heights above the ground at $z=$ 4, 8, 20, and 40 m. The data from these four sonic anemometers were sampled continuously at 10-Hz frequency and divided into half hour intervals. The time-synchronization among the four levels was ensured with the help of GPS clocks, supplied by the manufacturer. Apart from that, five all-in-one weather sensors (Gill MaxiMet GMX600) were also installed on the tower at $z=$ 2, 6, 10, 30, and 49 m. These sensors measured the horizontal wind speed and direction, temperature, relative humidity, pressure, and rainfall. The data from all-in-one weather sensors were logged into a CR3000 data-logger (Campbell Scientific Inc.) at both 1-min and 30-min intervals. 

Apart from the tower observations, a secondary dataset from a C-band (5.625 GHz) polarimetric Doppler weather radar has also been used in this study to establish the spatial features associated with the passage of the wind-gust event. The radar was positioned on the roof of a four-storeyed building, situated at a distance of around 1.5 km towards the North from the location of the tower. It had a range of 250 km and we used an image in the Plan Position Indicator (PPI) scan at an elevation angle of 1.5$^{\circ}$ to describe the gust front event. The gust front signature could be identified due to the resolution of the radar in the near range of 50 km, as the scans were configured with 1 $\mu$s pulse-width corresponding to 150 m range resolution. We used the radar reflectivity, which is the sixth power of the droplet size,  within a radius of 50 km around Solapur. The 40 dBZ reflectivity corresponds to 10 mm hr$^{-1}$ rain rate, identifying the heavy rain echos. Our specific interest was to investigate the formation and propagation of the gust front in association with any deep convective clouds over the region. 

\section{Results and discussion}
\label{results}
We begin with discussing the general meteorological characteristics associated with the gust front, as detected from the Doppler C-band radar. Subsequently, we present the results related to the effect on the surface measurements as obtained from the sensors on the 50-m instrumented micrometeorological tower. Later, we show other relevant results from persistence and clustering analyses to indicate any difference in the statistical properties of the turbulent structures as the gust front moved past the tower location. Plausible physical interpretations are also provided during the course of our investigation.  

\subsection{Meteorological characteristics of the gust front}
\label{gust}
\begin{figure}[h]
\centering
\hspace*{-0.6in}
\includegraphics[width=1.3\textwidth]{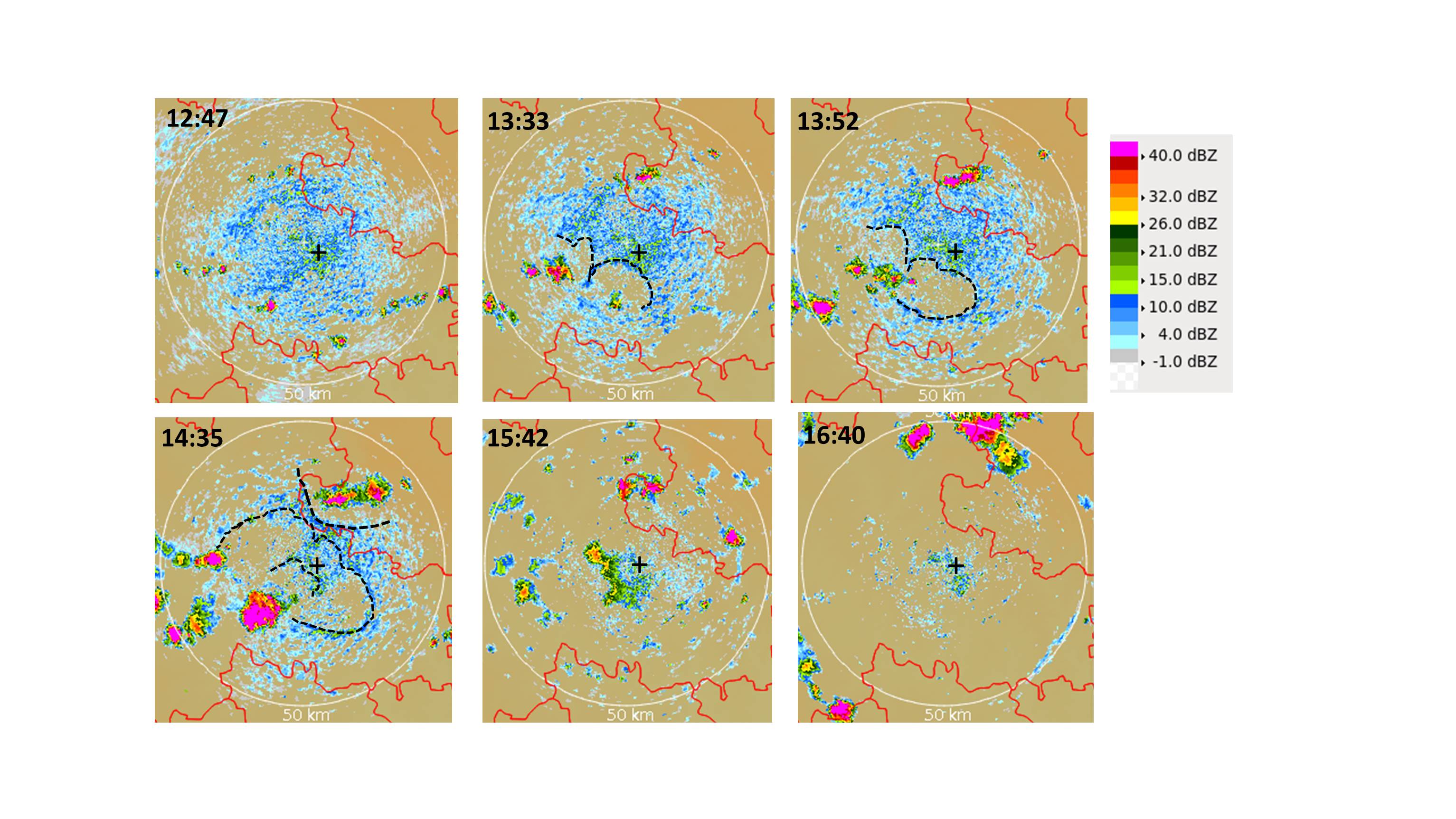}
  \caption{Radar reflectivities are shown as dBZ contours, obtained from a Doppler C-band radar at the lowest elevation within 50 km of Solapur, during the passage of a gust front on 22-September-2018. The colour bar on the right hand side of the panels show the reflectivity values in dBZ. The dotted lines indicate the observed gust front and the `+’ symbols specify the location of the 50-m micrometeorological tower. At the top left corner of all the panels, the local times are shown corresponding to each snapshot.}
\label{fig:2}
\end{figure}

Figure \ref{fig:2} shows the reflectivity from the C-band Doppler radar for a 50 km range at an elevation angle of 1.5$^{\circ}$, corresponding to six different PPI scans at six different local times of 22-September-2018 (GMT+05:30). Note that, each of the plots shown in Fig. \ref{fig:2} point towards the North. 

From the first panel of Fig. \ref{fig:2} (12:47 PM), we could notice the presence of few isolated convective cells at the South-West sector, identified as the contours of reflectivity greater than 30 dBZ. Additionally, several shallow non-precipitating cumulus clouds with reflectivity below 20 dBZ are also noted in and around the study area. However, as the time progressed (12:47 PM to 14:35 PM, first to fourth panels), these isolated convective cells formed wide clusters of deep-convective clouds commensurate with heavy precipitation (contour values more than 40 dBZ). Since the precipitation was noted over a wide area, this phenomenon cooled the surroundings and resulted in clear areas around, without clouds. This interpretation is supported by the observations, where we could see the majority of the areas around the deep convective clusters were mostly cloud-free on the South-West sector (see third and fourth panels). These clear areas were thus associated with the cold air which was pushing forward in the boundary layer and advanced towards the tower location (showed with a `+' sign in Fig. \ref{fig:2}). 

The boundaries between the clear and cloudy areas are designated with dashed lines in Fig. \ref{fig:2} (see the second, third, and fourth panels), which indicate the interface between the cold and the warm air. Apart from that, a convergence area could also be noted over the East and North-East sector of the tower location around 14:35 PM (see the fourth panel in Fig. \ref{fig:2}). Regardless, during the later times at 15:42 and 16:40 PM, the deep-convective clusters were not present in the proximity of the tower location (see the fifth and sixth panels in Fig. \ref{fig:2}). 

In summary, we interpret these observations from Fig. \ref{fig:2} as the evidence that the gust front, which originated from the outflows of several deep-convective clouds, approached the tower location starting from the period approximately around 14:00 PM. After the passage of the gust front, clear conditions existed over the tower location. Given the uniqueness of this event, it is an opportunity to investigate its associated effects on the surface measurements with a special focus on the consequent changes in the turbulent characteristics. We present results related to these aspects in the subsequent sections.  

\subsubsection{Surface measurements from all-in-one weather sensors}
\label{all-in-one}

\begin{figure}[h]
\centering
\hspace*{-1in}
\includegraphics[width=1.3\textwidth]{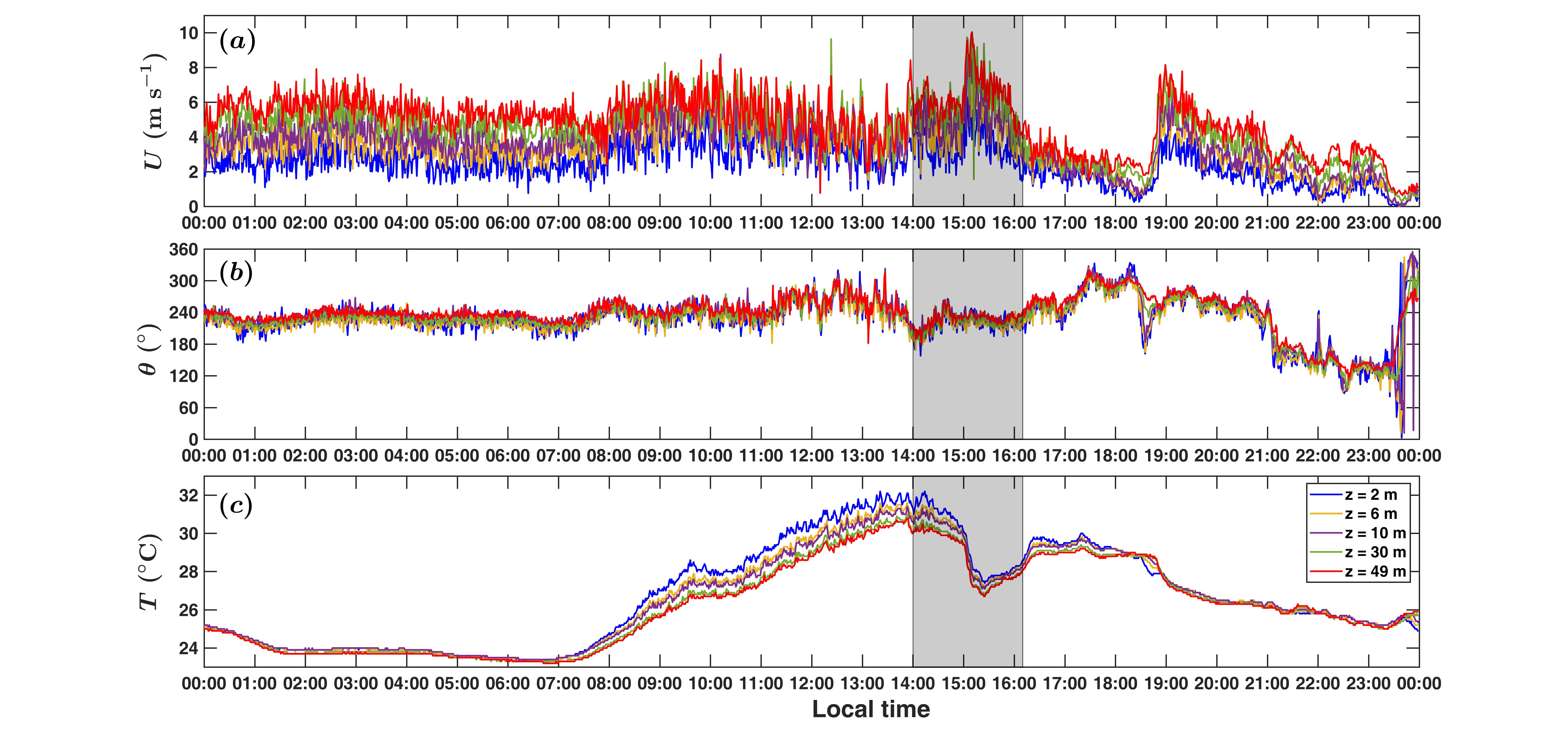}
  \caption{The 24-hr time series of every 1-min data from the five all-in-one weather sensors ($z=$ 2, 6, 10, 30, and 49 m) are shown for the variables such as: (a) horizontal wind speed (${U}$), (b) wind direction from the North (${\theta}$), and (c) dry-bulb temperature ($T$), corresponding to 22-September-2018. The grey shaded regions in all the panels indicate the period between 14:00-16:10 PM (local time), where a sudden decrease in the dry-bulb temperature is observed, with a near-constant wind direction from the South-West sector ($180^{\circ}<\theta<270^{\circ}$). The legend in panel (c) describe the different lines.}
\label{fig:3}
\end{figure}

Figure \ref{fig:3} shows the 24-hr time series of every 1-min data from the five all-in-one weather sensors ($z=$ 2, 6, 10, 30, and 49 m) corresponding to the variables such as: horizontal wind speed (${U}$), wind direction from the North (${\theta}$), and the dry-bulb temperature ($T$). By analysing the radar images from Fig. \ref{fig:2}, we inferred that a gust front approached the tower location starting from the period around 14:00 PM. To investigate its effect on the surface measurements, a grey shaded area is shown in Fig. \ref{fig:3} corresponding to a period between 14:00-16:10 PM (local time). 

One may notice from Figs. \ref{fig:3}a and c that, during this period there was an increase in the horizontal wind speed approximately beyond 15:00 PM, with a sudden drop around 4$^{\circ}$C in the dry-bulb temperature. Remarkably, this drop in $T$ happened across all the five heights on the tower, starting from $z=$ 2 m to $z=$ 49 m. However, before the occurrence of this event, the temperature increased with the time of the day as would be expected in a canonical convective surface layer as a result of surface heating. 

On the other hand, the wind-direction stayed approximately constant within the grey shaded region (14:00-16:10 PM) with the values between $180^{\circ}<\theta<270^{\circ}$, indicating the wind approached the tower from the South-West sector. This observation is consistent with the radar images, where we detected the movement of the gust front was from the South-West sector. Noticeably, beyond 16:10 PM the surge in the horizontal wind-speed subsided with values decreasing up to 4 m s$^{-1}$ at all the five heights. This decreasing trend in $U$ lasted up to 19:00 PM, after which an another surge was observed in the wind speed. 

In a nutshell, the above results imply that the gust front which passed the tower location, impacted the horizontal wind speed and dry-bulb temperature at all the five heights, continuing almost up to 16:10 PM. Moreover, it was fortuitous to note, the wind direction during this period stayed within the South-West sector, clear of any obstacles in the incoming flow (see Sect. \ref{Data}). Therefore, this allowed us to investigate the associated effect on the characteristics of wind speed and temperature, as obtained from the high-frequency sonic anemometer measurements.

\subsubsection{High-frequency measurements from the sonic anemometers}
\label{sonic_meas}

\begin{figure}[h]
\centering
\hspace*{-1in}
\includegraphics[width=1.3\textwidth]{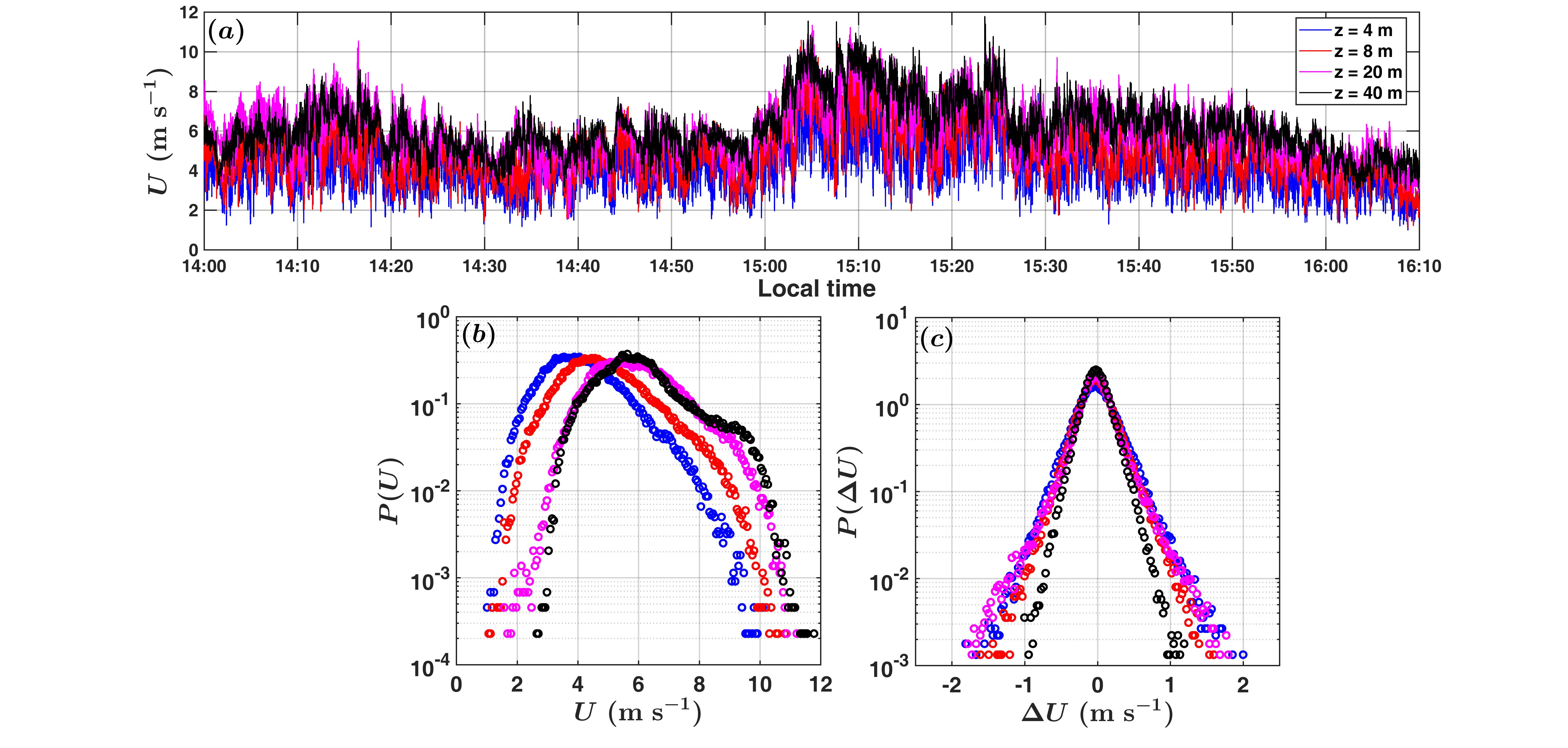}
  \caption{The 10-Hz time series of the horizontal wind speed are shown in panel (a) from the four sonic anemometers ($z=$ 4, 8, 20, and 40 m) corresponding to the period 14:00-16:10 PM. The bottom panels (b and c) show the respective probability density functions (PDF's) of the horizontal wind speed and acceleration during this period. The legend representing the colours associated with each height is shown in panel (a).}
\label{fig:4}
\end{figure}

Figure \ref{fig:4}a shows the time series of the horizontal wind speed, as obtained from the 10-Hz sonic anemometer measurements at four different heights, during the period 14:00-16:10 PM. In general, the high-frequency time series of $U$ presented in Fig. \ref{fig:4}a show similar characteristics as in Fig. \ref{fig:3}a, with a sudden increase being visible at times beyond 15:00 PM. Nevertheless, given the fine temporal resolution, the turbulent characteristics of $U$ such as their probability density functions (PDF's) can also be obtained from these measurements. 

Figures \ref{fig:4}b and c show the PDF's of $U$ and their temporal increments $\Delta U$ (acceleration), corresponding to the said period for all the four heights. A systematic shift towards the larger values is observed in the peak positions of the $U$ PDF's as the observation heights increased. Apart from that, there is also an indication of a secondary peak in the $U$ PDF's for the highest measurement level, i.e. at $z=$ 40 m. On the contrary, for the increment PDF's no such variation with height is observed in Fig. \ref{fig:4}b.  

Note that, the acceleration statistics are related to the small-scale (short-lived) eddies, given their association with the longitudinal gradient and hence with dissipation \citep{chu1996probability}. Therefore, the height-invariance of the acceleration PDF's as observed in Fig. \ref{fig:4}b, implies the variations of the wind speed in the surface layer during the gust front, were mainly dominated by the long-lived coherent eddies. Next we will present results for the temperature variations. 

\begin{figure}[h]
\centering
\hspace*{-1in}
\includegraphics[width=1.3\textwidth]{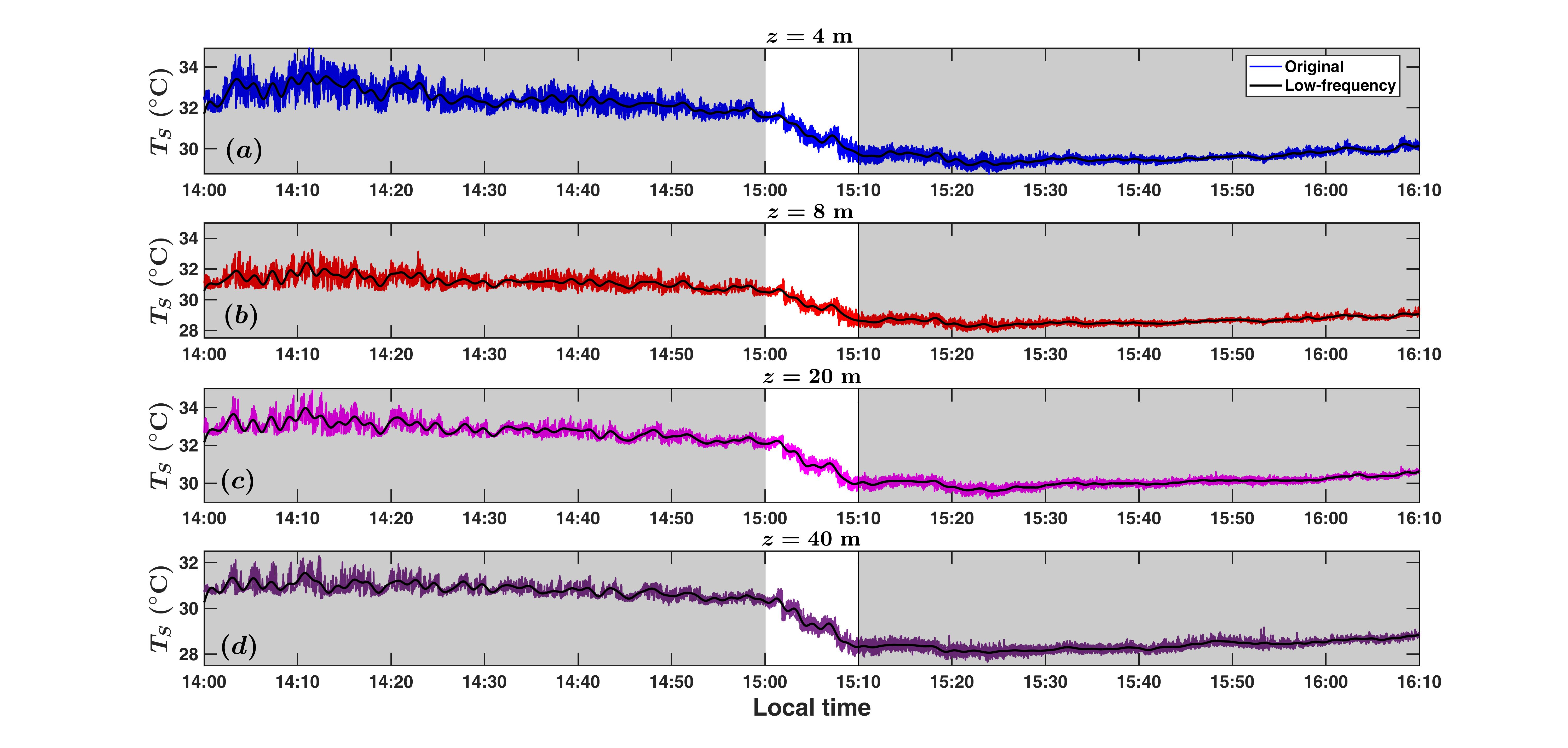}
  \caption{Same as in Fig. \ref{fig:4}, but for the sonic temperatures ($T_{S}$) shown for $z=$ (a) 4 m, (b) 8 m, (c) 20 m, and (d) 40 m. Similar to Fig. \ref{fig:3}c, a decrease in $T_{S}$ is visible between 15:00-15:10 PM at all the heights. The thick black lines on all the panels indicate a Fourier filtered low-frequency signal (threshold frequency set at 0.01 Hz) used to remove the trend from the time series to calculate the turbulent fluctuations. The grey shaded regions designate the two respective periods between 14:00-15:00 PM and 15:10-16:10 PM, which occurred before and after the drop in $T_{S}$.}
\label{fig:5}
\end{figure}

Figure \ref{fig:5} shows the high-frequency time series of the sonic temperatures ($T_{S}$) at four different heights, between the period 14:00-16:10 PM. The first thing what we notice from Fig. \ref{fig:5} is the sharp drop in $T_{S}$ between 15:00-15:10 PM. Physically we interpret this as, due to the passing of the gust front, resulted in a strong flow of cold air into the surface layer which caused such a drastic reduction in the temperature. It is incredible to note that, such reduction in temperature simultaneously existed at all the four measurement heights from $z=$ 4 to 40 m. 

Additionally, this drop in the temperature created an interface, which separated two different regimes. For their better identification, these two regimes are shaded in grey in Fig. \ref{fig:5}. One could notice, in the first regime (14:00-15:00 PM) the $T_{S}$ values exhibited significant temporal variation. Whereas in the second regime (15:10-16:10 PM), the intensity of the variation in $T_{S}$ was extremely weak. Apart from that, in the period between 14:00-15:00 PM, the $T_{S}$ time series displayed the classical ramp-cliff patterns, commonly observed in convective surface layers. This was in sharp contrast with the period between 15:10-16:10 PM, where such signatures were completely absent.

Undoubtedly, this type of a phenomenon presents an interesting case to investigate the respective turbulent characteristics of the temperature variations associated with the periods before and after the drop in the temperature. In fact, to gain a deeper perspective, one could ask, \emph{whether there is any systematic difference between the structural properties of the turbulent temperature fluctuations related to their size distributions and temporal organization?}. However, since this is a non-trivial occurrence, the computation of the turbulent temperature fluctuations are difficult given the variation in the mean state itself. 

Therefore, in order to compute the turbulent fluctuations the apparent trend in the $T_{S}$ time series need to be removed. To accomplish that, we computed the Fourier spectrum of the $T_{S}$ time series for the whole period (14:00-16:10 PM). We noted that, for frequencies lesser than or equal to 0.01 Hz, the Fourier amplitudes displayed a clear linear trend. From \citet{kaimal1994atmospheric}, we know such trend is associated with low-frequency oscillations in a turbulent time series which need to be removed in order to compute the fluctuations. We thus applied a Fourier filter where the contributions from the frequencies lesser than 0.01 Hz were removed and then the inverse transformation was applied to retrieve the filtered time series. 

The thick black lines shown in all the four panels of Fig. \ref{fig:5} indicate this Fourier-filtered low-frequency variation. To compute the turbulent fluctuations in temperature ($T^{\prime}$), this trend was removed from the $T_{S}$ series. We next investigate the respective structural properties of the turbulent $T^{\prime}$ signals for the periods between 14:00-15:00 PM and 15:10-16:10 PM.  

\subsection{The turbulent structures of temperature variations}
\label{temp_var}
We commence our description of structural properties of turbulence associated with the two periods (shown as the grey shaded regions of Fig. \ref{fig:5}), by discussing the differences in the PDF's of the temperature fluctuations and their temporal increments. This is because, the tails of the PDF's of any turbulent fluctuations are influenced by the coherent eddy motions, whereas the tails of the increment PDF's are governed by the small-scale eddy motions \citep{chu1996probability,pouransari2015statistical}.

\subsubsection{PDF's of temperature fluctuations and increments}
\label{temp_pdfs}

\begin{figure}[h]
\centering
\hspace*{-1.3in}
\includegraphics[width=1.4\textwidth]{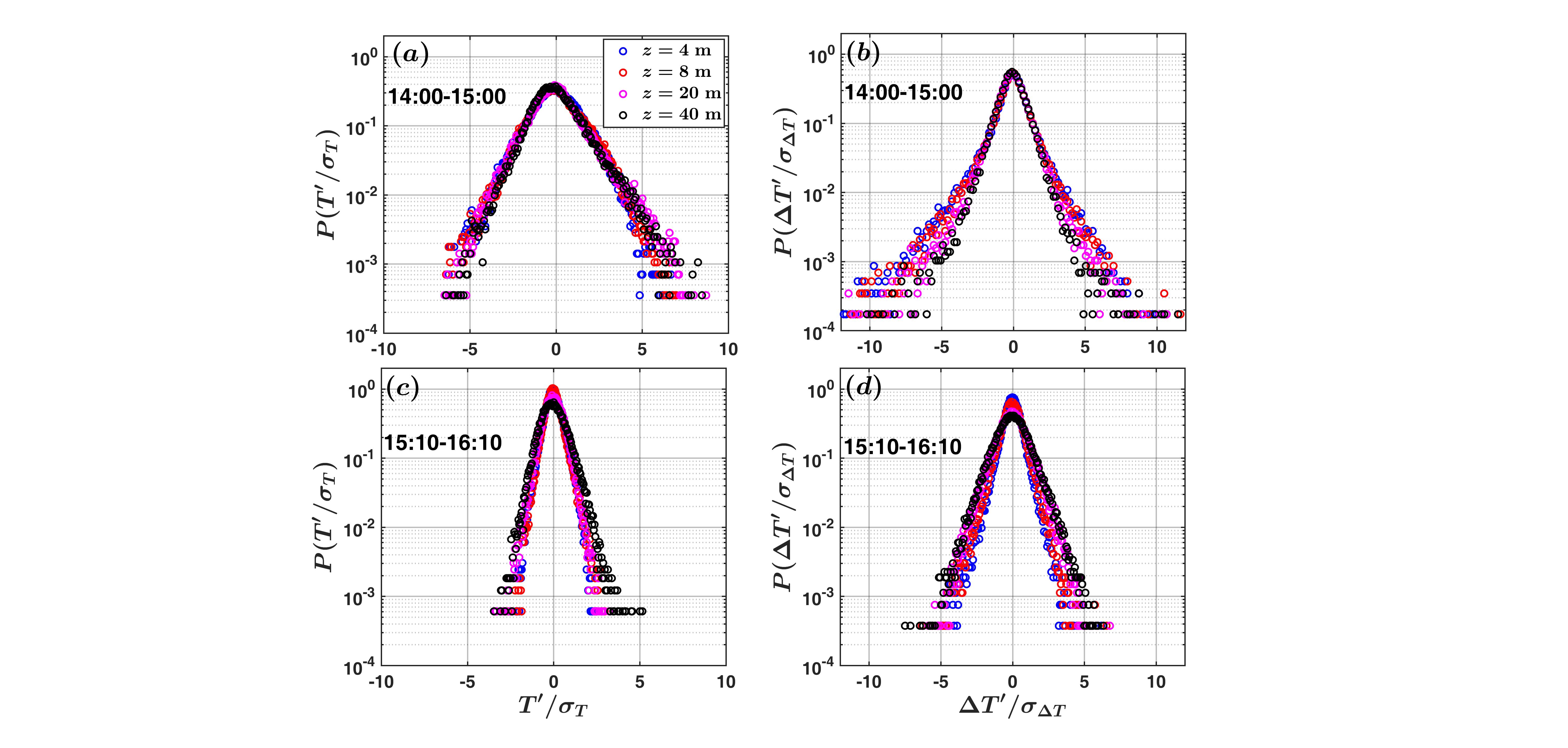}
  \caption{The PDF's associated with turbulent temperature fluctuations ($T^{\prime}/\sigma_{T}$) and their increments ($\Delta T^{\prime}/\sigma_{\Delta T}$) are shown separately for the two periods as designated in Fig. \ref{fig:5}. The top panels (a) and (b) show the respective PDF's corresponding to the fluctuations and their increments during the period 14:00-15:00 PM. Similar information are also presented in the bottom panels (c) and (d), but for the period 15:10-16:10 PM. Note that, the fluctuations and increments are normalized by their respective standard deviations, computed over the entire period 14:00-16:10 PM, after removal of the low-frequency trend from the sonic temperature signals. The legend representing the colours associated with each height is shown in panel (a).}
\label{fig:6}
\end{figure}

Figure \ref{fig:6} shows the PDF's of the temperature fluctuations ($T^{\prime}$) and its increments ($\Delta T^{\prime}$) for the periods 14:00-15:00 PM and 15:10-16:10 PM (see Fig. \ref{fig:5}). Note that, the fluctuations and the increments in Fig. \ref{fig:6} are normalized by their respective standard deviations ($\sigma_{T}$ and $\sigma_{\Delta T}$) computed over the whole 14:00-16:10 PM period, after the removal of the low-frequency trend. From Figs. \ref{fig:6}a and c, we notice that the $T^{\prime}$ PDF's for both the periods are approximately symmetric with respect to 0 and do not reveal any substantial difference with height. Nevertheless, the $T^{\prime}$ PDF's for the period 14:00-15:00 PM exhibit a significantly broader range in fluctuations compared to the period 15:10-16:10 PM. This observation is consistent with our visual inspection of Fig. \ref{fig:5}, where the temperature signals at all the heights are apparently more jagged in the first period as opposed to the next.

Apart from that, from Figs. \ref{fig:6}b and d we also observe a substantial difference in the increment PDF's between the two periods. For all the four heights, the increment PDF's in Fig. \ref{fig:6}b display a heavy left tail associated with the large negative values and an extruded peak at the smaller values. This feature is remarkably consistent with \citet{chu1996probability}, where they also observed similar characteristics in an unstable surface layer. They explained this heavy left tail as a consequence of the abundance of ramp-cliff patterns present in the temperature signal during the convective conditions. However, in Fig. \ref{fig:6}d no such evidence of a heavy tail could be found in the increment PDF's corresponding to all the four heights. Contrarily, the increment PDF's in Fig. \ref{fig:6}d look strikingly similar to the PDF's of the fluctuations itself in Fig. \ref{fig:6}b.

To summarize, the aforementioned results imply that during the period 14:00-15:00 PM the statistical characteristics of the large- and small-scale turbulent structures are considerably different, given the clear discrepancy in the shapes of the fluctuation and increment PDF's. On the other hand, for the period 15:10-16:10 PM, the fluctuation and increment PDF's display very similar character, with almost no change being observed in their shapes. This points out a statistical equivalence between the large- and small-scale structures when the temperature fluctuations are weak and remain quiescent. Therefore, one may ask, \emph{how this difference in the statistical attributes between the energetic and the quiescent periods is related to the size distributions and clustering tendencies of the turbulent structures}?

Note that, from the above discussed PDF's no information can be obtained about the time or length scales (sizes) of the associated turbulent structures. This is because, while performing the binning exercise to construct the PDF's we mask any dependence with time. Nonetheless, in a turbulent signal the positive and negative fluctuations occur with a range of different time scales as they tend to exit and re-enter to their respective states \citep{kalmar2019complexity,chowdhuri2019revisiting}. To extract that additional information, we turn our attention to persistence analysis. 

\subsubsection{Persistence analysis of temperature structures}
\label{persistence}
Persistence is defined as the probability that the local value of a fluctuating field does not change its sign for a certain amount of time \citep{majumdar1999persistence}. Put differently, the concept of persistence is also equivalent to the distributions of the inter-arrival times between the successive zero-crossings of a stochastic signal \citep{chamecki2013persistence}. The zero-crossings in a stochastic signal are identified by using a telegraphic approximation (TA) of its fluctuations ($x^{\prime}$), expressed as,
\begin{equation}
(x^{\prime})_{\rm TA}=\frac{1}{2}(\frac{x^{\prime}(t)}{|x^{\prime}(t)|}+1),
\label{TA}
\end{equation}
and locating the points where this TA series changes its values from 0 (off state) to 1 (on state) or vice-versa. For laboratory boundary layer flows, \citet{sreenivasan1983zero} and \citet{kailasnath1993zero} interpreted the distributions of the persistence time scales as the size distributions of the turbulent structures, after converting those to spatial length scales by employing Taylor's frozen turbulence hypothesis. For further details on the application of persistence in turbulent flows, the readers may consult \citet{chowdhuri2020persistence}.

\begin{figure}[h]
\centering
\hspace*{-1.3in}
\includegraphics[width=1.4\textwidth]{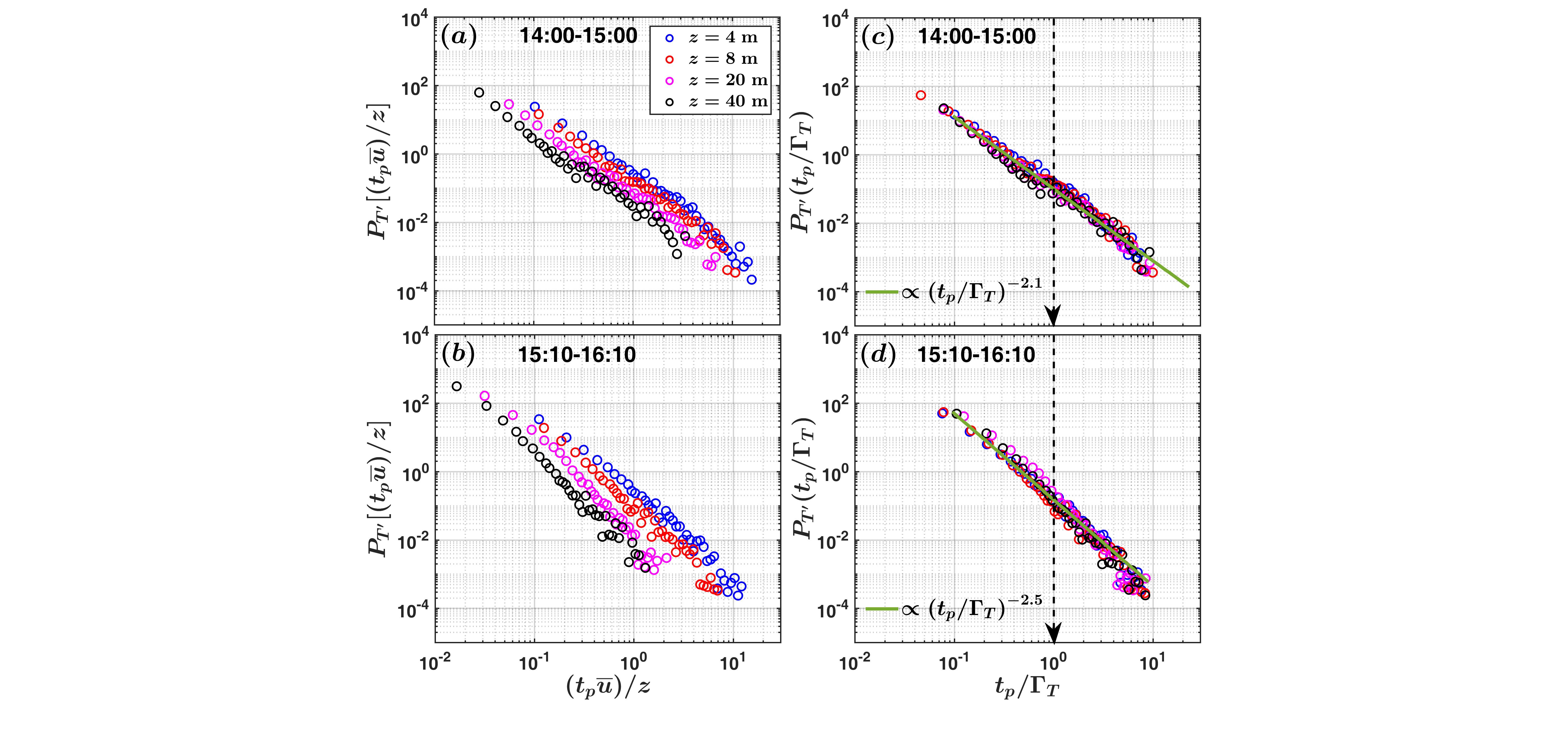}
  \caption{The persistence PDF's of the turbulent temperature fluctuations ($T^{\prime}$) are plotted separately for the same two periods as shown in Fig. \ref{fig:5}. The persistence time scales ($t_{p}$) are normalized by two different ways such as: they are converted to a streamwise size $(t_{p}\overline{u})$ ($\overline{u}$ is the mean horizontal wind speed) and normalized by $z$ (see panels (a) and (b)) or they are normalized directly by the integral scale $\Gamma_{T}$ associated with temperature (see panels (c) and (d)). Due to the second normalization, a good collapse is observed in the persistence PDF's for all the heights with a clear power-law signature. The associated power-law functions are shown in the legends of panels (c) and (d). The legend representing the colours associated with each height is shown in panel (a). The dashed black lines with solid arrows in panels (c) and (d) indicate the position $t_{p}/\Gamma_{T}=$ 1.}
\label{fig:7}
\end{figure}

Figure \ref{fig:7} shows the persistence PDF's or equivalently the size distributions of the turbulent structures associated with the temperature fluctuations, corresponding to the same two periods as indicated in Fig. \ref{fig:5}. Note that, a log-log representation is chosen to display the PDF's, such that any power-law functions in such plots would be shown as straight lines. To compute these PDF's, we follow the same methodology as detailed in \citet{chowdhuri2020persistence}. 

Figures \ref{fig:7}a and b show the size distributions for the two periods, where the persistence time ($t_{p}$) is converted to a streamwise length scale ($t_{p}\overline{u}$, where $\overline{u}$ is the mean horizontal wind speed as obtained from Fig. \ref{fig:4}) and normalized by the height $z$ above the surface. This normalization is chosen under the assumption that in a convective surface layer the turbulent structures are self-similar with height \citep{kader1991spectra}. In spite of that, one may notice, normalizing the persistence length scales with $z$ does not collapse the size distributions as there is a clear separation among the PDF's at different heights (Figs. \ref{fig:7}a and b). 

To investigate this further, in Figs. \ref{fig:7}c and d the same information are presented, but the $t_{p}$ values are normalized by the integral time scales of temperature fluctuations. We use such normalization because, \citet{chowdhuri2020persistence} demonstrated that the statistical characteristics of the persistence PDF's are related to the temporal coherence of the time series, expressed by its integral scale. These integral time scales ($\Gamma_{T}$) are computed separately for the two periods, from the first zero-crossings of the auto-correlation functions of the temperature fluctuations \citep{kaimal1994atmospheric,katul1997energy,li2012mean}. In Fig. \ref{fig:s1} of the supplementary material, we provide the respective auto-correlation functions and the integral scales for these two periods and their vertical variation corresponding to different measurement levels.  

We note that, by applying this particular normalization with the integral length scale, the size distributions of the turbulent structures collapsed for all the heights, following a power-law distribution. The exponents of these power-laws were computed by performing a linear regression on the log-log plots. We obtained the best fit values ($R^{2}>0.96$) for the exponents (slopes of the straight lines) as $2.1$ and $2.5$ respectively for the two corresponding periods (Figs. \ref{fig:7}c and d). The difference in these exponents between the two periods remains unexplained at present. 

Notwithstanding the above limitation, it is intriguing to note that the power-law feature of the persistence PDF's in Figs. \ref{fig:7}c and d extend for the time scales even larger than the integral scale. This observation is in disagreement with the results from canonical convective surface layers. \citet{cava2012role} and \citet{chowdhuri2020persistence} have demonstrated that in a statistically stationary convective surface layer for a wide range of stability conditions (spanning from highly-convective to near-neutral), the persistence PDF's display a power-law distribution for time scales smaller than the integral scales. They attributed this to the self-similar Richardson cascading mechanism, commonly observed in a well-developed turbulent flow. In addition to that, they also noticed at scales larger than the integral scales, there was an exponential drop in the persistence PDF's, hallmark of a Poisson type process. \citet{cava2012role} surmised this phenomenon as a consequence of random deformation of the large coherent structures, giving rise to several sub-structures with independent arrival times. 

However, for the present context such an exponential drop is not visible in Figs. \ref{fig:7}c and d. Since power-laws are synonymous with scale-invariance \citep{newman2005power,verma2006universal}, it implies the entire size distributions of the turbulent structures for both the periods (14:00-15:00 PM and 15:10-16:10 PM) are scale-free in nature. This indicates the passing of the gust front initiated a scale-free response which governed the turbulent characteristics of the temperature fluctuations, generating a self-similar size distribution of the associated structures. 

It is noteworthy to mention that this type of a situation has qualitative similarities with the self organized criticality (SOC) observed in the sandpile model of \citet{bak1988self}, famously known as the BTW experiment. \citet{lewis2010cause} and \citet{accikalin2017concept} have specified that the SOC occurs in such complex systems which operate at a point near to the critical state, where a small external perturbation can create avalanches having a power-law size distribution. To provide such a heuristic analogy with the present case in hand, we expect the deep-convective cells whose outflows generated the gust front, acted as an external stimuli which disturbed its surroundings beyond the tipping point and created a scale-free response. Over the course of time, this response propagated to the surface layer of the convective boundary layer and generated structures having self-similar size distributions. 

Despite sound promising, at present the aforementioned connection with SOC is a hypothesis and future research in this direction is required to corroborate it further. Be that as it may, we next investigate whether these self-similar temperature structures displayed any clustering tendency related to their temporal organization during these two periods.  

\subsubsection{Clustering properties of temperature structures}
\label{clustering}
While the persistence analysis describes the size distributions of the turbulent structures, the clustering or aggregation property is related to the temporal organization pattern of these structures. From a statistical point of view, a time series is clustered if the on and off states of the corresponding TA series (see Eq. \ref{TA}) are distributed unevenly in time \citep{cava2009effects,poggi2009flume,cava2012role,cava2019submeso}. 

\begin{figure}[h]
\centering
\hspace*{-1.6in}
\includegraphics[width=1.6\textwidth]{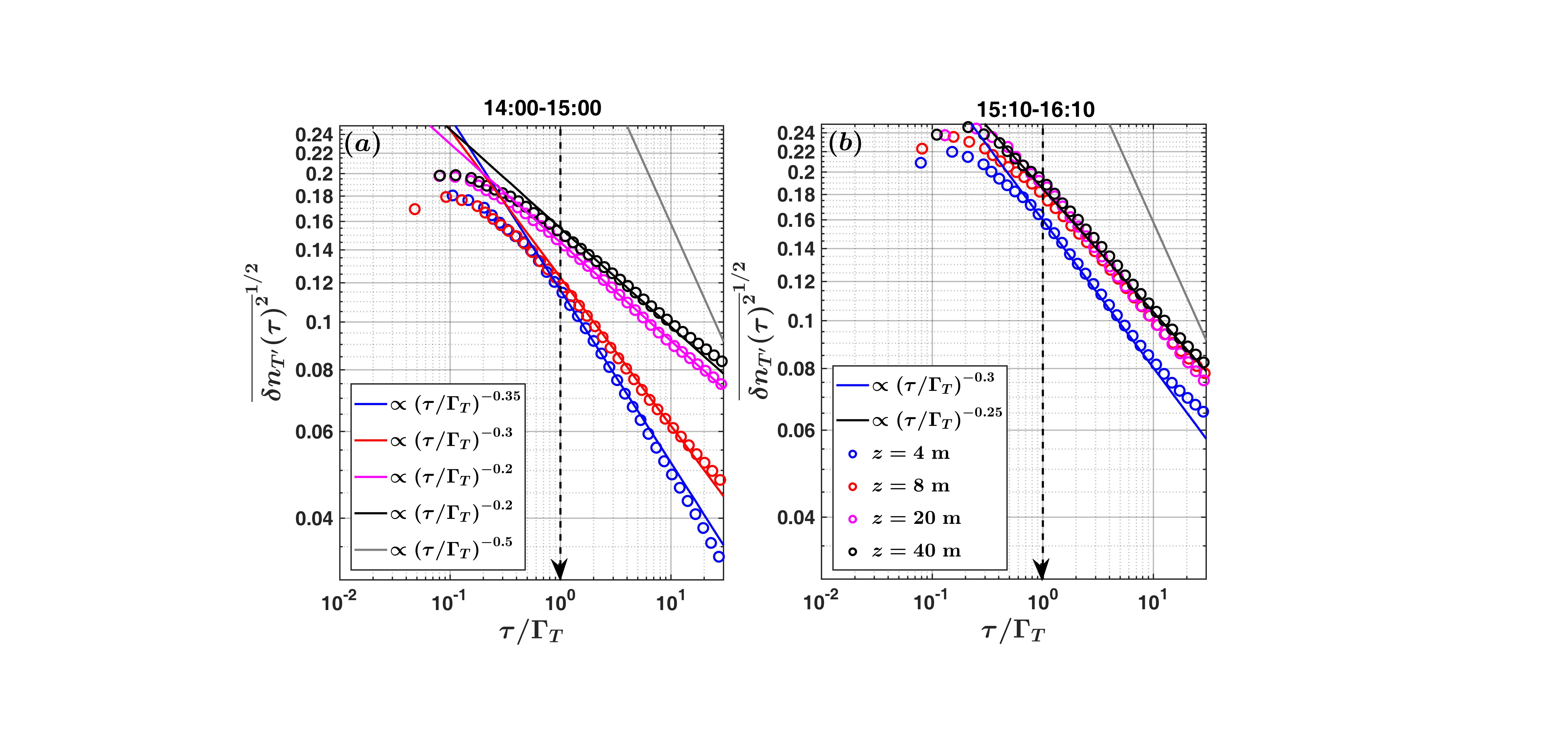}
  \caption{The RMS values of the zero-crossing density fluctuations $\Big[{\overline{{{\delta n}_{T^{\prime}}(\tau)}^{2}}}^{1/2}\Big]$ are shown for the $T^{\prime}$ signals, plotted against the normalized lags $\tau/\Gamma_{T}$, during the periods (a) 14:00-15:00 PM and (b) 15:10-16:10 PM. The clustering exponents are computed from the power-law fits to the RMS zero-crossing density fluctuations at different lags, as shown in the legends of panels (a) and (b). The grey lines in both the panels denote the power-law exponent of $0.5$, corresponding to a white-noise signal which displays no clustering effect. The dashed black lines with solid arrows indicate the position $\tau/\Gamma_{T}=$ 1.}
\label{fig:8}
\end{figure}

To quantify such behaviour, a clustering exponent ($\alpha$) is computed such as,
\begin{equation}
{\overline{{{\delta n}(\tau)}^{2}}}^{1/2} \propto \tau^{-\alpha},
\label{CE}    
\end{equation}
where $\tau$ are the time lags and ${\overline{{{\delta n}(\tau)}^{2}}}^{1/2}$is the root-mean-square (RMS) of the zero-crossing density fluctuations, defined as,
\begin{equation}
{\delta n}(\tau)=n_{\tau}(t)-\overline{n_{\tau}(t)},
\label{CE_1}    
\end{equation}
where $n_{\tau}(t)$ represents the zero-crossing densities at each $\tau$. A step-by-step implementation of the method to compute the clustering exponents is provided by \citet{poggi2009flume}, which we have followed in this study. Additionally, \citet{sreenivasan2006clustering} showed that for a white noise signal which displays no clustering tendency, the clustering exponent (see Eq. \ref{CE}) equals to 0.5. Thus, if $\alpha<0.5$, it is an indication that the turbulent structures have a tendency to aggregate or cluster \citep{poggi2009flume}.  

Figure \ref{fig:8} shows the RMS values of the zero-crossing density fluctuations for the $T^{\prime}$ signals during the two periods (14:00-15:00 PM and 15:10-16:10 PM), corresponding to all the four heights. Note that, the time lags ($\tau$) shown in Fig. \ref{fig:8} are normalized by the integral scales of temperature ($\Gamma_{T}$), associated with the two periods. The clustering exponents ($\alpha$) are computed by fitting power-laws to these RMS values at different lags, as shown in Fig. \ref{fig:8}. Similar to Fig. \ref{fig:7}, a log-log representation is chosen so the power-laws are shown as straight lines. Moreover, for comparison purpose, the grey lines on both the panels indicate the clustering exponent of 0.5 related to a white noise signal.  

From Fig. \ref{fig:8}a, we note that for the period 14:00-15:00 PM, the clustering exponents of the $T^{\prime}$ signals were significantly different from 0.5 ($\alpha<0.5$) at all the four heights. Apart from that, a clustering tendency was observed for the turbulent structures greater than the sizes of the integral scales. Nevertheless, for the lower two levels ($z=$ 4 and 8 m), a slight break in the slopes of the clustering exponents was visible at the scales approximately equal to the integral scales. However, as the heights increased ($z=$ 20 and 40 m), this difference almost disappeared and a substantial drop was noted in the clustering exponents ($\alpha \approx$ 0.2) compared to the lower two levels ($\alpha \approx$ 0.3). On the other hand, for the period 15:10-16:10 PM, the clustering exponents corresponding to all the four heights were approximately equal to each other, as could be seen from Fig. \ref{fig:8}b ($\alpha \approx$ 0.3). This indicates, even though the turbulent structures are self-similar for both the periods, their clustering properties with height were significantly different. 

Despite such discrepancy with Fig. \ref{fig:8}a, a similar clustering tendency was observed in Fig. \ref{fig:8}b at the scales larger than the integral scales, with $\alpha$ values being substantially smaller than 0.5. This observation is in clear variance with the case presented by \citet{cava2012role}. They found that, in a canonical convective surface layer, at scales larger than the integral scales, the turbulent structures displayed no clustering tendency as the exponents ($\alpha$) approached 0.5. Therefore, an important question arises, \emph{whether such disparity is related to the self-similar nature of the turbulent temperature structures extending over all the scales of motions as a result of a scale-free response associated with the gust front?}. A plausible hypothesis could be, the passage of the gust front in a convective surface layer generates a cascading effect which permeates across all the scales and modulates the aggregation properties of the turbulent structures even at those sizes larger than the integral scale. However, the further confirmation of this scale-modulation effect is beyond the scope of the present article and reserved for our future endeavours. We present our conclusions in the next section.

\section{Conclusion}
\label{concl}
In this paper, we present a case study utilising the simultaneous observations from a C-band Doppler weather radar and an instrumented micrometeorological tower with multi-level measurements, to delineate the influence of a gust front on the surface layer turbulence in a tropical convective boundary layer. We find that, for this particular case, a gust front detected from the Doppler weather radar passed over the location of the 50-m micrometeorological tower. This gust front was originated from the outflows of several deep-convective clouds, placed within the 50 km radius of the tower location. Due to the intrusion of the cold air associated with the gust front, a drop in the temperature was noted at heights within the surface layer. We investigated the consequent effects of this on the turbulent temperature characteristics and the following results emerged: 
\begin{enumerate}
    \item Due to the passage of a gust front, a sudden drop of $\approx 4^{\circ}$C was noted in the temperature at heights within the surface layer. Additionally, it was found that this drop in the temperature created an interface which separated two different regimes. In one regime, the temperature fluctuations were large and energetic, whereas in the other regime they were weak and quiescent.
    \item By investigating the structural properties of the turbulent temperature fluctuations associated with these two regimes, we discovered that the size distributions of the turbulent structures for both of these regimes displayed a clear power-law signature. Since power-laws are synonymous with scale-invariance, this indicated the passing of the gust front initiated a scale-free response which governed the turbulent characteristics of the temperature fluctuations. To explain this, we provide a heuristic analogy with self organized criticality (SOC) as observed in the complex systems.  
    \item Despite the self-similar nature of the turbulent structures, their aggregation or clustering properties differed between these two regimes. For the regime corresponding to large temperature fluctuations, the turbulent structures were significantly clustered, whose clustering properties changed with height. However, for the second regime where the temperature fluctuations were weak, there was comparatively a lesser tendency to cluster with no discernible change being observed with height.
\end{enumerate}

In summary, going back to the two research questions raised in the introduction regarding 1) the modification of the clustering due to the presence of gust front and 2) the scale related effect, we show that there is a definite clustering tendency for the turbulent temperature structures at scales larger than the integral scales. This finding is at odds with a canonical convective surface layer, where the turbulent structures display no clustering tendency at scales larger than the integral scales. A plausible reason for this oddity is, in our case with the presence of a gust front, the self-similar nature of the temperature turbulent structures extend for scales even larger than the integral scales. 

Last but not the least, in general the clustering or aggregation properties are related to small-scale phenomena, which cause anomalous scaling in the inertial subrange of the turbulence spectrum. In the present study, we observe during the existence of a gust front, the turbulent temperature structures exhibit clustering at scales larger than the integral scales. Since integral scales are associated with energy-containing motions, this preliminary evidence suggests a scale-modulation effect where the small scales influence the larger scales. Therefore, for our future research on the effect of gust front in convective surface layer turbulence, it would be worth to ask,

\emph{Do the vertical velocity fluctuations also display similar characteristics and what are the associated implications on the turbulent flux modelling?}

\section*{Data availability}
On reasonable request, the datasets analysed during the current study can be made available to the interested researchers by contacting Thara V Prabha (\href{mailto:thara@tropmet.res.in}{thara@tropmet.res.in}). The computer codes needed to reproduce the figures are available by contacting the corresponding author Subharthi Chowdhuri at \href{mailto:subharthi.cat@tropmet.res.in}{subharthi.cat@tropmet.res.in}. 

\section*{Conflict of Interest}
The authors declare that they have no conflict of interest.

\section*{Author contributions}
The authors Subharthi Chowdhuri and Thara V Prabha conceptualized the study. The data collection was performed by Subharthi Chowdhuri, Kiran Todekar, Anand K Karipot, and Palani Murugavel. All the analyses for the paper were carried out by Subharthi Chowdhuri. The first draft of the manuscript was written by Subharthi Chowdhuri and all authors commented on previous versions of the manuscript. All authors read and approved the final manuscript.

\section*{Acknowledgements}
Cloud Aerosol Interaction and Precipitation Enhancement Experiment is conducted by the Indian Institute of Tropical Meteorology, which is an autonomous institute and fully funded by the Ministry of Earth Sciences, Government of India. Authors are grateful to several colleagues who contributed to the success of the CAIPEEX project. The authors also acknowledge the local support and hospitality provided by N. B. Navale Sinhgad College of Engineering (NBNSCOE), Kegaon-Solapur, during the experiment. The author Subharthi Chowdhuri expresses his gratitude to Dr. Tirtha Banerjee and to Dr. Tam\'{a}s Kalm\'{a}r-Nagy for many fruitful discussions on the concepts of persistence, zero-crossing densities, and SOC phenomenon. 

\bibliographystyle{apalike}  
\bibliography{references} 
\clearpage

\section*{Supplementary material}
\renewcommand{\thefigure}{S\arabic{figure}}
\renewcommand{\figurename}{Fig.}
\setcounter{figure}{0}

\begin{figure}[h]
\centering
\vspace*{0.5in}
\hspace*{-1.6in}
\includegraphics[width=1.6\textwidth]{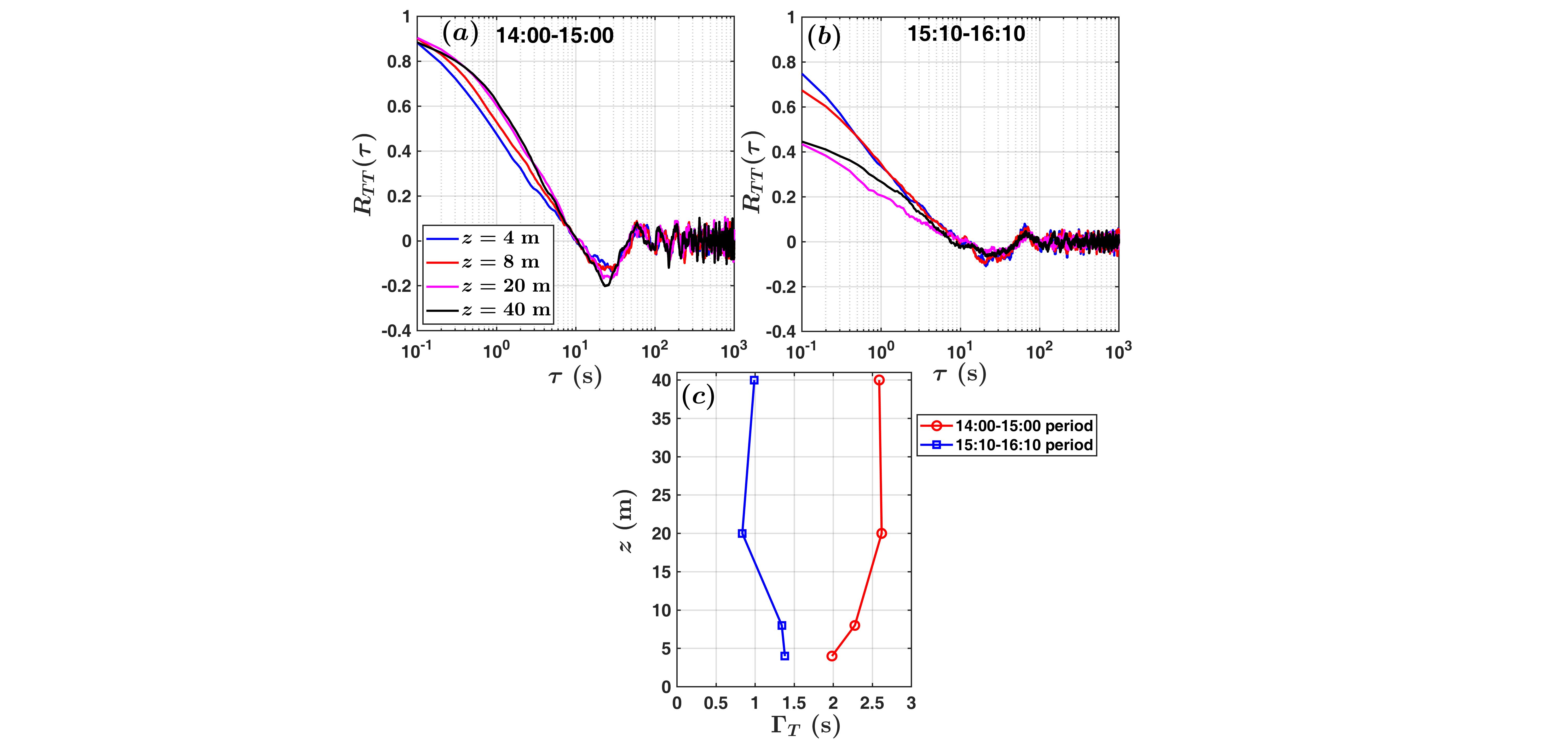}
 \caption{The top panels show the auto-correlation functions ($R_{TT}(\tau)$) of the turbulent temperature fluctuations plotted against the lags ($\tau$), for the periods (a) 14:00-15:00 PM and (b) 15:10-16:10 PM, corresponding to all the four heights. Note that, given the log axis in the abscissa for panels (a) and (b), the $\tau$ values start from 0.1 s (sampling interval) instead of 0. The bottom panel displays the vertical profiles of the integral time scales ($\Gamma_{T}$), computed from the integration of the auto-correlation functions up to the first zero-crossing. The legends in panels (a) and (c) describe the different markers.}
\label{fig:s1}
\end{figure}

\end{document}